\documentclass[12pt,draftcls,onecolumn]{IEEEtran}
\usepackage{amsmath,amsfonts}
\usepackage{algorithmic}
\usepackage{array}
\usepackage[caption=false,font=normalsize,labelfont=sf,textfont=sf]{subfig}
\usepackage{textcomp}
\usepackage{stfloats}
\usepackage{url}
\usepackage{verbatim}
\usepackage{graphicx}
\def\BibTeX{{\rm B\kern-.05em{\sc i\kern-.025em b}\kern-.08em
    T\kern-.1667em\lower.7ex\hbox{E}\kern-.125emX}}
\usepackage{balance}
\usepackage{physics}

\begin{document}
\bibliographystyle{IEEEtran}    

\title{Dyadic-Order Quantum Fractional Transforms: Circuit Constructions and Applications to Hartley and Cosine Transform Families}

\author{
Matheus J. A. Oliveira, 
Israel F. Araujo, 
José R. de Oliveira Neto,
Juliano B. Lima
\thanks{
This work has been supported in part by Conselho Nacional de Desenvolvimento Científico e Tecnológico (CNPq) under grants 422306/2023-1, 312935/2023-4, and 405903/2023-5, and Financiadora de Estudos e Projetos (MCTI/FINEP/FNDCT/CENTROS TEMÁTICOS 2023 Grant No 1020/24).
}
}

\maketitle

\begin{abstract}
This paper presents a generalized circuit framework for constructing Shih-type fractionalizations of unitary operators of dyadic order, i.e., operators $U$ satisfying $U^{2^n}=I$. 
Building upon the architecture of the quantum fractional Fourier transform (QFrFT), we show that fractionalization can be implemented coherently as a weighted superposition of integer powers,
$\sum_k c_k(\alpha)U^k$,
where the coefficients are generated through an ancilla-domain quantum Fourier transform and a diagonal phase modulation. 
Under the assumption that controlled implementations of the required powers of $U$ are available, the resulting circuit yields a parameterized family of operators that interpolates the integer powers of $U$ and satisfies the additive property of fractional transforms. 
As concrete applications, we derive explicit quantum circuit realizations of the quantum fractional Hartley transform (QFrHT) and of the fractional cosine-transform families associated with Types~I and~IV. 
These constructions demonstrate the versatility of the proposed dyadic-order fractionalization framework for structured operators arising in quantum signal processing.
\end{abstract}



\section{Introduction} \label{intro}

Quantum computing leverages the principles of quantum mechanics to provide computational advantages for specific classes of problems~\cite{nielsen2000quantum}. 
While the conceptual foundation was laid by Feynman in 1982~\cite{Feynman1982}, the field gained widespread attention after Shor demonstrated polynomial-time integer factorization on a quantum computer~\cite{SHOR94}. 
The exponential speedup in Shor's algorithm relies on the quantum Fourier transform (QFT), underscoring the central role of signal-processing primitives in quantum algorithms.


In recent years, significant progress has been made in quantum signal processing (QSP)~\cite{Martyn2025parallelquantum} and related areas such as quantum neural networks~\cite{QCNN2022}, quantum image representation~\cite{FRQI2011,NEQR2013}, and quantum watermarking~\cite{QWM2019}.
Given the ubiquity of the QFT, several classical signal-processing transforms have been adapted to the quantum domain, such as the quantum Hartley transform (QHT) and the quantum cosine transforms (QCTs)~\cite{Klappenecker:qct:qht:2001}.

In classical signal processing, fractional transforms provide a parameterized interpolation between integer powers of periodic operators and have been widely studied in applications ranging from filtering to optics~\cite{frft_eulogio,frft_wavelets,frft_grafos}. 
The quantum fractional Fourier transform (QFrFT) was only recently introduced~\cite{Zhao2023}. 
However, a unified circuit-level methodology for fractionalizing other structured quantum operators is still lacking.


In this work, we focus on unitary operators of dyadic order, that is, operators $U$ satisfying $U^{2^n}=I$. 
This structural condition is fundamental rather than incidental: it implies that the eigenvalues of $U$ are dyadic roots of unity, enabling exact binary-phase encoding within an $n$-qubit ancilla register. 
Exploiting this property, we show that fractionalization can be implemented coherently as a weighted superposition of integer powers,
\[
\sum_{k=0}^{2^n-1} c_k(\alpha) U^k,
\]
where the coefficients are generated via an ancilla-domain quantum Fourier transform followed by a diagonal phase modulation.

Building upon the architecture of the QFrFT, we derive a generalized circuit construction applicable to dyadic-order unitary operators. 
As concrete instantiations, we present explicit circuit realizations for the quantum fractional Hartley transform (QFrHT) and the fractional cosine/sine transform structures associated with Types~I and~IV. 
These constructions demonstrate the versatility of the proposed dyadic-order fractionalization framework for structured operators arising in quantum signal processing.


The remainder of this paper is organized as follows. 
Section~\ref{sec:preliminares} reviews the notation and quantum circuit primitives used throughout the work. 
Section~\ref{sec:qfrft} revisits the QFrFT circuit and highlights the structural mechanism underlying its fractionalization procedure. 
Section~\ref{sec:new_transforms} presents the generalized dyadic-order fractionalization framework and derives the circuit implementations for the QFrHT and the fractional cosine/sine transform structures associated with Types~I and~IV. 
Finally, Section~\ref{sec:conclusion} summarizes the main results and outlines directions for future research.

\section{Preliminaries}\label{sec:preliminares}

\subsection{Qubits and Vector Representation}
\label{subsec:qubits}

A qubit is represented by a normalized vector in a two-dimensional complex Hilbert space ($\mathcal{H}^2$).
Using the standard Dirac notation, the computational basis states are defined as column vectors: %
$\ket{0} = [1,0]^T$ and $\ket{1} = [0,1]^T$.
A general single-qubit state $\ket{\psi}$ is a linear superposition of these basis states, expressed as
\begin{equation*}
    \ket{\psi} = \alpha \ket{0} + \beta \ket{1} = \begin{bmatrix} \alpha \\ \beta \end{bmatrix},
\end{equation*}
where $\alpha, \beta \in \mathbb{C}$ are probability amplitudes satisfying the normalization condition $|\alpha|^2 + |\beta|^2 = 1$.

For a system composed of $n$ qubits, the state space expands to a $2^n$-dimensional Hilbert space via the tensor product ($\otimes$).
A generic basis state in this system is denoted as $\ket{u} = \ket{u_{n-1}} \otimes \dots \otimes \ket{u_0}$, where $u_i \in \{0,1\}$. 

\subsection{Quantum Gates and Matrix Representation} \label{subsec:gate}

Quantum algorithms are described by quantum circuits composed of unitary gates acting on qubits. 
The fundamental single-qubit operators used throughout this work are the Pauli gates ($X$, $Y$, $Z$), the Hadamard gate ($H$), and the phase gate $P(\phi)$, given by
\begin{equation*}
\begin{split}
X &=
\begin{bmatrix}
0 & 1\\
1 & 0
\end{bmatrix},
\qquad
Y =
\begin{bmatrix}
0 & -i\\
i & 0
\end{bmatrix},
\qquad
Z =
\begin{bmatrix}
1 & 0\\
0 & -1
\end{bmatrix},\\[0.4em]
H &= \frac{1}{\sqrt{2}}
\begin{bmatrix}
1 & 1\\
1 & -1
\end{bmatrix},
\qquad
P(\phi)=
\begin{bmatrix}
1 & 0\\
0 & e^{i\phi}
\end{bmatrix}.
\end{split}
\end{equation*}

In this work, we adopt the standard circuit model in which an $n$-qubit gate is represented by a $2^n\times 2^n$ unitary matrix. 
A subscript on an operator indicates the number of qubits on which it acts. For example,
\begin{equation*} \label{eq:exemplo_notation}
    H_2 = H\otimes H
    = \frac{1}{2}
    \begin{bmatrix}
        1 & 1 & 1 & 1\\
        1 & -1 & 1 & -1\\
        1 & 1 & -1 & -1\\
        1 & -1 & -1 & 1
    \end{bmatrix}.
\end{equation*}


\subsection{Controlled Gates and Quantum Circuits} \label{subsec:circuits}

A critical mechanism for 
implementing conditional logic in quantum algorithms is the controlled operation.
A generic controlled-unitary gate, denoted by $CU$, acts on at least two qubits: a control qubit and a target system.
Mathematically, the operation applies the unitary $U$ to the target if and only if the control qubit is in the state $\ket{1}$; if the control is in $\ket{0}$, the target remains unchanged.
The matrix representation of a $CU$ gate in the standard basis $\{\ket{00}, \ket{01}, \ket{10}, \ket{11}\}$ is given by the block-diagonal matrix
\begin{equation*}
    CU = \begin{bmatrix} I & 0 \\ 0 & U \end{bmatrix},
\end{equation*}
where $I$ is the $2 \times 2$ identity matrix.

\begin{figure}[h]
\begin{center}
\includegraphics[width=8.4cm]{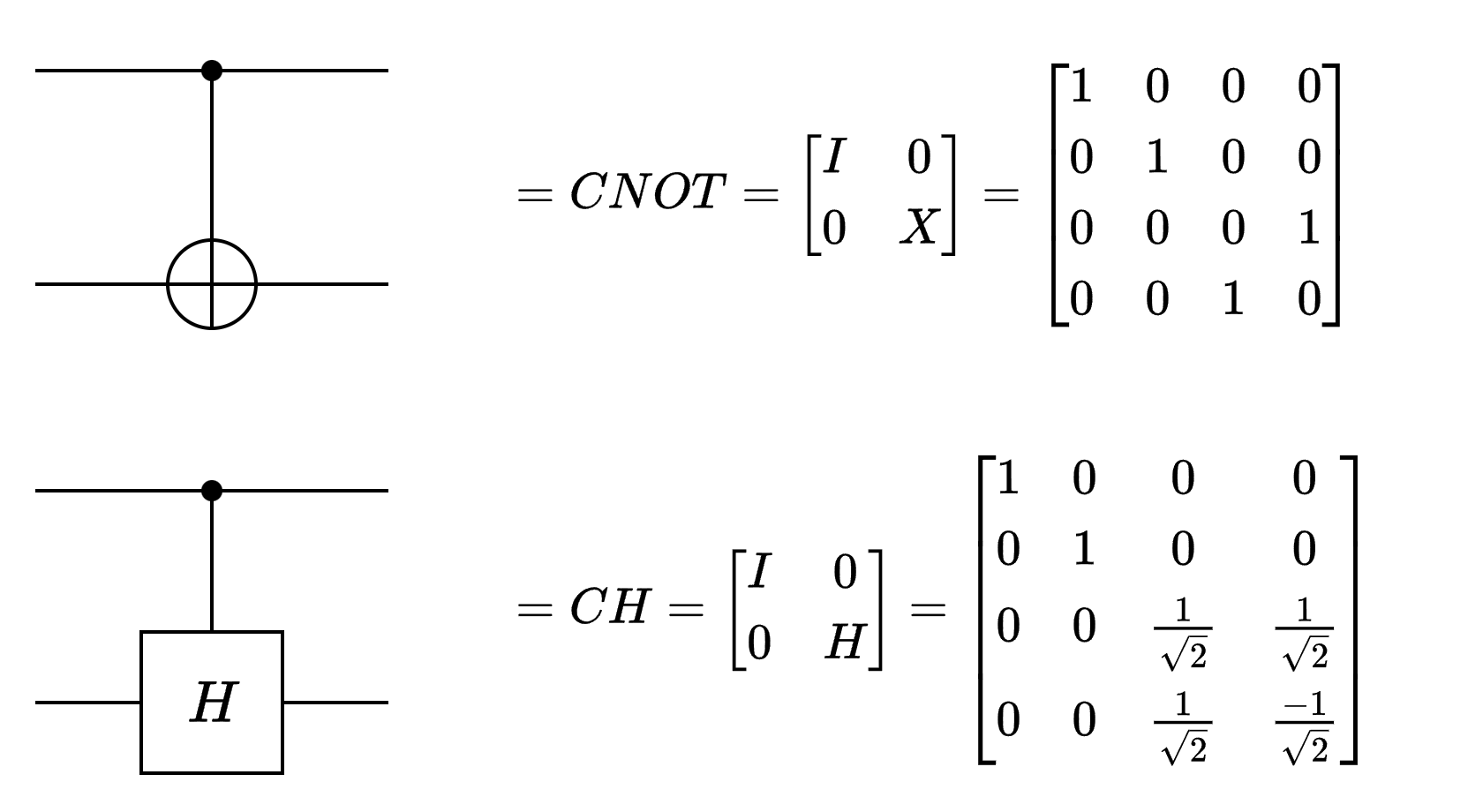}    
\caption{The controlled-NOT gate (CNOT) at the top and a controlled-Hadamard gate (CH) at the bottom.}
\label{fig:circuito_exemplo}
\end{center}
\end{figure}

Quantum algorithms are commonly represented using standard quantum circuit diagrams, as illustrated in Figure~\ref{fig:circuito_exemplo}.
In this graphical representation, time flows from left to right, and each horizontal line corresponds to a qubit or a register of qubits.
By convention, the top-most line represents the most significant qubit (MSB) of the register.
Figure~\ref{fig:circuito_exemplo} depicts examples of a controlled-NOT (CNOT) and a controlled-Hadamard gate, illustrating this notation.

Further details on the fundamentals of quantum computation can be found in~\cite{nielsen2000quantum}.

\subsection{Operator Order and Fractionalization} \label{subsec:op_order_frac}

Let $U$ be a unitary operator. 
The operator is said to have order $N$ if

\[
U^N = I.
\]

In classical signal processing, several transforms possess finite order. 
For instance, the discrete Fourier transform satisfies $F^4 = I$, while the Hartley and cosine transforms are involutions satisfying $U^2 = I$.

Fractionalization defines a parameterized family of operators $U^\alpha$, with $\alpha \in \mathbb{R}$, that interpolates the integer powers of $U$. 
A consistent fractionalization must satisfy the additivity property

\[
U^\alpha U^\beta = U^{\alpha+\beta}.
\]

For convenience, we introduce the block-diagonal operator

\[
D_N(U) = \mathrm{diag}(I, U, U^2, \dots, U^{N-1}),
\]
which will be used to represent multiplexed powers of a unitary operator within quantum circuits.

In this work, we focus on operators of dyadic order, that is, unitary operators satisfying

\[
U^{2^n} = I .
\]

Such operators have spectra consisting of $2^n$-th roots of unity. 
This property enables exact binary-phase encoding using an $n$-qubit ancilla register, which forms the basis for the fractionalization framework developed in the following sections.


\section{Review of the Quantum Fractional Fourier Transform} \label{sec:qfrft}

\begin{figure}[h]
\begin{center}
\includegraphics[width=13.4cm]{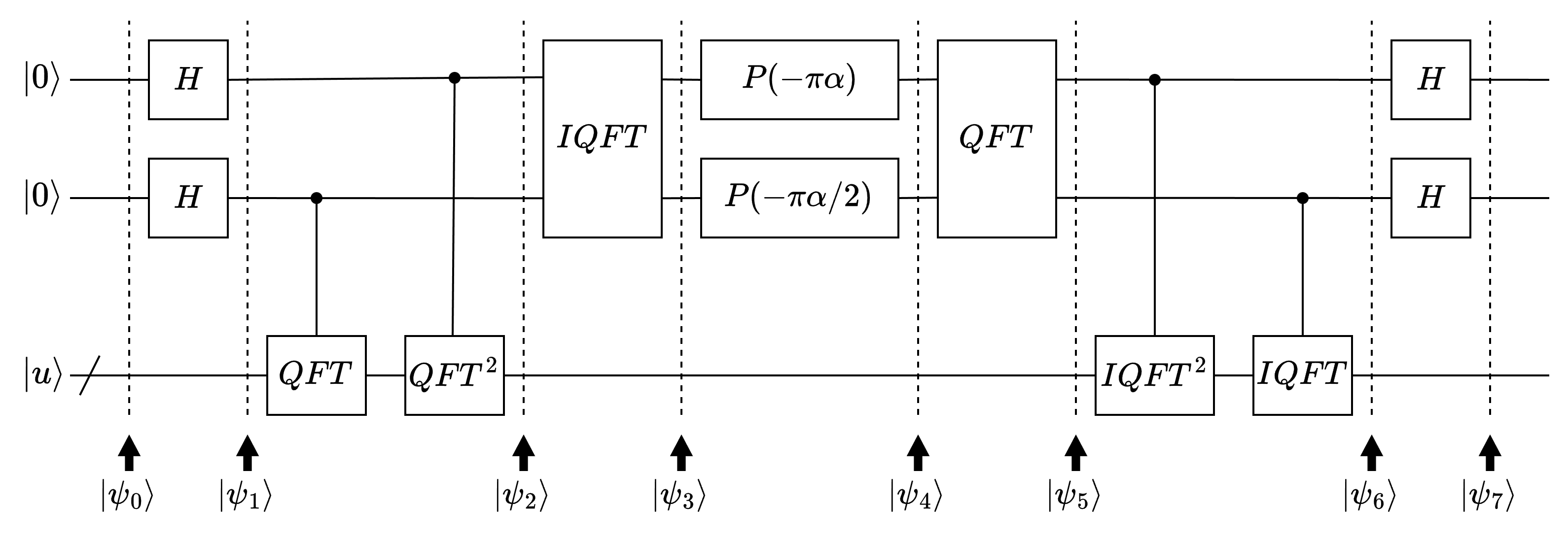}    
\caption{Quantum circuit for the QFrFT.} 
\label{fig:circuito_qfrft}
\end{center}
\end{figure}

This section revisits the quantum fractional Fourier transform (QFrFT) circuit proposed in~\cite{Zhao2023} and reformulates its derivation in a way that highlights the structural mechanism underlying the construction. 
Starting from the circuit shown in Figure~\ref{fig:circuito_qfrft}, we derive the operator implemented on the input state and show that the QFrFT can be expressed as a coherent linear combination of the integer powers of the Fourier operator. 
This viewpoint will serve as the basis for the generalized dyadic-order fractionalization framework developed in the next section.

The circuit uses two ancilla qubits, initialized in the state $\ket{00}$, which control the application of different powers of the Fourier transform.
Let $\ket{u}$ denote the input quantum state upon which the fractional Fourier transform acts, consisting of $q$ qubits.
Consequently, the total system requires a register of size $q+2$ qubits.
The initial state is 
\begin{equation*} \label{eq:ket0}
    |\psi_0\rangle = |00\rangle|u\rangle.
\end{equation*}

The first operation prepares the ancilla register in an equal superposition, yielding
\begin{equation*} \label{eq:ket1}
\begin{split}
    |\psi_1\rangle = \Big(H_2 \otimes I_{q}\Big)|\psi_0\rangle 
    = \frac{1}{2}(|00\rangle + |01\rangle + |10\rangle + |11\rangle)|u\rangle .
\end{split}
\end{equation*}

The next block applies, conditioned on the two ancilla qubits, the integer powers of the Fourier operator acting on the $q$-qubit data register. 
Using the multiplexed notation introduced in Section~\ref{sec:preliminares}, this operation is written as
\[
D_4(F)=\mathrm{diag}(I,F,F^2,F^3),
\]
where $F\equiv F_q$ denotes the $2^q$-point quantum Fourier transform.
Accordingly, the next state is
\begin{equation*} \label{eq:ket2}
\begin{split}
     |\psi_2\rangle = D_4(F)
     |\psi_1\rangle
=    \frac{1}{2}\big(|00\rangle I|u\rangle + |01\rangle F|u\rangle +|10\rangle F^2|u\rangle + |11\rangle F^3|u\rangle\big).
\end{split}
\end{equation*}

The inverse Fourier transform acting on the two ancilla qubits is the 4-point inverse Fourier matrix,
\begin{equation*} \label{eq:iqft_matriz}
\begin{split}
    F^{-1}_{2} &=\frac{1}{2} 
\begin{bmatrix} 
1 & 1 & 1 & 1 \\ 
1 & w^{-1} & w^{-2} & w^{-3} \\ 
1 & w^{-2} & 1 & w^{-2} \\ 
1 & w^{-3} & w^{-2} & w^{-1}
\end{bmatrix},
\end{split}
\end{equation*}
where $w = e^{-i\frac{2\pi}{4}}$. 
Thus,
\begin{equation*} \label{eq:psi_3_1}
\begin{split}
     |\psi_3\rangle = &\Big( F^{-1}_{2} \otimes I_{q} \Big)|\psi_2\rangle, \\
= &\frac{1}{4} \Big[ 
   |00\rangle (  I + F + F^2 + F^3)|u\rangle 
  +|01\rangle ( I + w^{-1}  F + w^{-2}  F^2 + w^{-3}  F^3)|u\rangle + \\
  &+|10\rangle ( I + w^{-2}  F + F^2 + w^{-2} F^3)|u\rangle  
  +|11\rangle ( I + w^{-3}  F + w^{-2}  F^2 + w^{-1}  F^3)|u\rangle \Big].
\end{split}
\end{equation*}


From $|\psi_3\rangle$ to $|\psi_4\rangle$, the ancilla register undergoes a diagonal phase modulation,
\begin{equation*}
    |\psi_4\rangle = \Big( P(2\alpha\theta_0) \otimes P(\alpha\theta_0) \otimes I_q \Big)|\psi_3\rangle,
\end{equation*}
where $\theta_0=-\frac{2\pi}{4}$ and $P(\phi)=\mathrm{diag}(1,e^{i\phi})$.
Equivalently, defining $w=e^{-i2\pi/4}$, this block can be written as
\begin{equation*}
\begin{split}
P(2\alpha\theta_0)\otimes P(\alpha\theta_0)
=
\begin{bmatrix}
1 & 0 & 0 & 0\\
0 & w^\alpha & 0 & 0\\
0 & 0 & w^{2\alpha} & 0\\
0 & 0 & 0 & w^{3\alpha}
\end{bmatrix}
= D_4(w^\alpha),
\end{split}
\end{equation*}
with the convention $w^\alpha := e^{-i\frac{2\pi}{4}\alpha}$.
The negative sign in the phase exponent is chosen to match the Fourier kernel convention used in the matrices $F_2$ and $F_2^{-1}$.


Accordingly, the phase-modulated state is
\begin{equation*} \label{eq:psi_4_3} 
\begin{split}
|\psi_4\rangle = \frac{1}{4} \Big[ 
    &|00\rangle ( I + F + F^2 + F^3)|u\rangle  \\
  +& |01\rangle ( w^{\alpha}I + w^{\alpha} w^{-1} F + w^{\alpha} w^{-2}  F^2 + w^{\alpha} w^{-3}  F^3)|u\rangle \\
  +& |10\rangle (w^{2\alpha} I + w^{2\alpha} w^{-2} F + w^{2\alpha} F^2 + w^{2\alpha} w^{-2} F^3)|u\rangle  \\
  +& |11\rangle (w^{3\alpha} I + w^{3\alpha} w^{-3} F + w^{3\alpha} w^{-2}  F^2 + w^{3\alpha} w^{-1}  F^3)|u\rangle \Big]
\end{split}
\end{equation*}


Applying the ancilla Fourier transform $F_2$ to $|\psi_4\rangle$, we obtain
\begin{equation*} \label{eq:psi_5_1}
     |\psi_5\rangle = \Big( F_2 \otimes I_{q} \Big)|\psi_4\rangle ,
\end{equation*}
where
\begin{equation*} \label{eq:qft_2}
    F_{2} =\frac{1}{2}  
\begin{bmatrix} 
1 & 1 & 1 & 1 \\ 
1 & w^{1} & w^{2} & w^{3} \\ 
1 & w^{2} & 1 & w^{2} \\ 
1 & w^{3} & w^{2} & w^{1}
\end{bmatrix}.
\end{equation*}

This yields the state 
\begin{equation*} \label{eq:psi_5_2} 
\begin{split}
|\psi_5\rangle = \frac{1}{2} \Big[ 
  |00\rangle \Big(& 
     \langle v_{\alpha},\bar{v}_{0} \rangle I 
    +\langle v_{\alpha},\bar{v}_{1} \rangle F 
    +\langle v_{\alpha},\bar{v}_{2} \rangle F^2 
    +\langle v_{\alpha},\bar{v}_{3} \rangle F^3 
  \Big) |u\rangle \\
  +|01\rangle \Big(& 
     \langle v_{\alpha},\bar{v}_{3} \rangle I 
    +\langle v_{\alpha},\bar{v}_{0} \rangle F 
    +\langle v_{\alpha},\bar{v}_{1} \rangle F^2 
    +\langle v_{\alpha},\bar{v}_{2} \rangle F^3
  \Big) |u\rangle \\
  +|10\rangle \Big(& 
     \langle v_{\alpha},\bar{v}_{2} \rangle I 
    +\langle v_{\alpha},\bar{v}_{3} \rangle F 
    +\langle v_{\alpha},\bar{v}_{0} \rangle F^2 
    +\langle v_{\alpha},\bar{v}_{1} \rangle F^3
  \Big) |u\rangle  \\
  +|11\rangle \Big(& 
     \langle v_{\alpha},\bar{v}_{1} \rangle I 
    +\langle v_{\alpha},\bar{v}_{2} \rangle F 
    +\langle v_{\alpha},\bar{v}_{3} \rangle F^2 
    +\langle v_{\alpha},\bar{v}_{0} \rangle F^3
  \Big) |u\rangle \Big],
\end{split}
\end{equation*}
where
\[
v_{\alpha}=\frac{1}{2}[1,w^{\alpha},w^{2\alpha},w^{3\alpha}],
\qquad
\bar v_k=\frac{1}{2}[1,w^{-k},w^{-2k},w^{-3k}],
\quad k=0,1,2,3,
\]
and $\langle v_\alpha,\bar v_k\rangle$ denotes the inner product between these vectors.
Equivalently, the coefficients can be written as
\[
\langle v_\alpha,\bar v_k\rangle
=
\frac{1}{4}\sum_{m=0}^3 w^{m(\alpha-k)},
\]
which makes explicit that the circuit generates interpolation weights over the integer powers of $F$.


Applying the inverse multiplexed Fourier powers to the data register, the state evolves as
\begin{equation*} \label{eq:psi_6_1}
|\psi_6\rangle = 
D_4(F^{-1})
|\psi_5\rangle.
\end{equation*}

Using the periodicity of the Fourier operator, namely $F^4=I$, so that $F^{-1}=F^3$, $F^{-2}=F^2$, and $F^{-3}=F$, the expression simplifies to
\begin{equation*} \label{eq:psi_6_22}
|\psi_6\rangle = \frac{1}{2} \Big(
   |00\rangle + |01\rangle + |10\rangle +|11\rangle \Big)
   \Big(
     \sum_{k=0}^3 \langle v_{\alpha} , \bar{v}_{k} \rangle F^{k}
  \Big) |u\rangle.
\end{equation*}


Applying a final Hadamard transform to the ancilla register yields
\begin{equation*} \label{eq:psi_7_1}
\begin{split}
    |\psi_7 \rangle =& \Big(H_2 \otimes I_{q}\Big)|\psi_6\rangle, \\
=&   
  |00\rangle \Big( 
  \sum_{k=0}^3 \langle v_{\alpha} , \bar{v}_{k} \rangle F^{k}
 \Big) |u\rangle,
\end{split}
\end{equation*}
Hence, the ancilla register is restored deterministically to the state $|00\rangle$, while the data register undergoes the operator
\[
FrFT_q(\alpha)=\sum_{k=0}^3 \langle v_\alpha,\bar v_k\rangle F^k,
\]
which is precisely the Shih fractional Fourier transform~\cite{SHIH95}. 
The QFrFT circuit therefore realizes the fractional operator as a coherent linear combination of the integer powers of the Fourier transform, with coefficients generated by the ancilla-domain Fourier-processing block.

In compact form, for a parameter $\alpha\in\mathbb{R}$, the circuit of Figure~\ref{fig:circuito_qfrft} implements
\begin{equation*} \label{eq:qfrft_complete}
\begin{split}
QFrFT_{q+2} \big(\alpha \big) \{|00u\rangle\} =& 
\Biggl( 
\Big(H_2 \otimes I_{q}\Big) 
D_4(F^{-1}) 
\Big( 
F_2 D_4(w^{\alpha}) F^{-1}_2 \otimes I_{q} \Big) \times \\
& \times D_4(F)
\Big(H_2 \otimes I_{q}\Big) \Biggl) |00u\rangle \\ 
=& |00\rangle\otimes FrFT_q(\alpha)\{|u\rangle\}.
\end{split}
\end{equation*}
This compact operator expression makes clear that the QFrFT construction is built from four ingredients: ancilla superposition, multiplexed powers of $F$, diagonal phase modulation, and ancilla uncomputation. 
These same ingredients will be generalized in the next section into a circuit framework for unitary operators of dyadic order.


\section{Quantum Fractional Transforms}\label{sec:new_transforms}

\subsection{Fractionalization method}\label{sec4sub1}

Shih showed that the construction underlying the classical fractional Fourier transform can be extended to operators of finite order~\cite{SHIH95}. 
In the present quantum setting, we consider unitary operators $U$ satisfying
\[
U^N=I,
\]
and adopt an equivalent formulation based on inner products rather than explicit trigonometric expansions. 
This representation is more compact and makes the interpolation structure of the fractional operator more transparent.

For an operator of order $N$, the Shih-type fractionalization is written as
\begin{equation*}
    FrU(\alpha)=\sum_{k=0}^{N-1}\langle v_\alpha,\bar v_k\rangle\,U^k,
\end{equation*}
where
\[
v_{\alpha}=\frac{1}{\sqrt{N}}[1,w^{\alpha},\dots,w^{(N-1)\alpha}],
\qquad
\bar v_k=\frac{1}{\sqrt{N}}[1,w^{-k},\dots,w^{-(N-1)k}],
\]
for $k=0,\dots,N-1$, with $w=e^{-i\frac{2\pi}{N}}$, and $\langle v_\alpha,\bar v_k\rangle$ denoting the inner product between the vectors.
Equivalently,
\[
\langle v_\alpha,\bar v_k\rangle
=
\frac{1}{N}\sum_{m=0}^{N-1} w^{m(\alpha-k)}.
\]

\begin{figure}[h]
\begin{center}
\includegraphics[width=16.0cm]{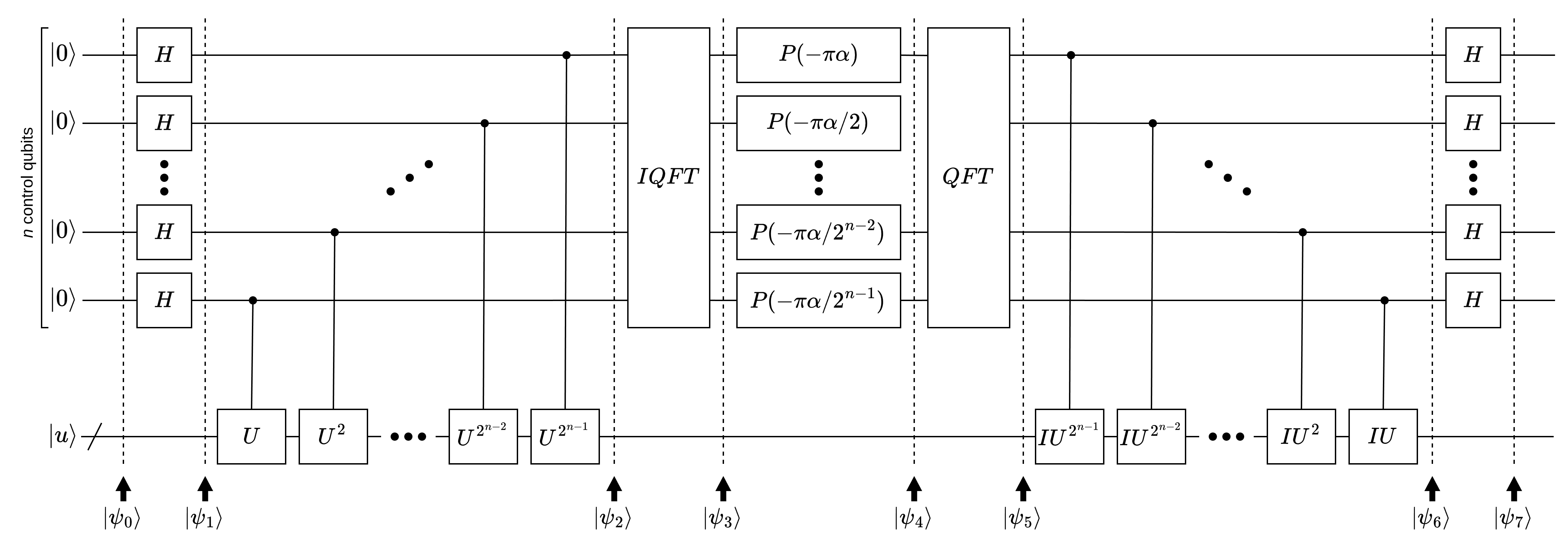}    
\caption{Diagram representing the quantum circuit for the generalized fractional transform.} 
\label{fig:circuito_qgfrt}
\end{center}
\end{figure}

For a quantum circuit implementation, the most natural case arises when the order of the operator is dyadic, i.e.,
\[
U^{2^n}=I,
\]
since an $n$-qubit ancilla register can then encode the corresponding phase indices exactly in binary form. 
Building on the circuit architecture reviewed in Section~\ref{sec:qfrft}, we obtain a generalized construction for unitary operators of dyadic order, provided that the required multiplexed powers of $U$ can be implemented. 
The resulting circuit is shown in Figure~\ref{fig:circuito_qgfrt}.

Repeating the same circuit logic used in Section~\ref{sec:qfrft}, the generalized circuit of Figure~\ref{fig:circuito_qgfrt} implements
\begin{equation*} \label{eq:qfrft_complete}
\begin{split}
QFrU_{q+n} \big(\alpha \big) \{|0\dots0u\rangle\} =& 
\Biggl( 
\Big(H_n \otimes I_{q}\Big) 
D_{2^n}(U_q^{-1})
\Big( 
F_n D_{2^n}(w^{\alpha}) F^{-1}_n \otimes I_{q} \Big) \times \\
& \times D_{2^n}(U_q)
\Big(H_n \otimes I_{q}\Big) \Biggl) |0\dots0u\rangle \\ 
=& |0\dots0\rangle_n \otimes FrU_q(\alpha)\{|u\rangle\} \\
=& |0\dots0\rangle_n \Big( 
 \sum_{k=0}^{2^n-1}  \langle v_{\alpha} , \bar{v}_{k} \rangle U^{k} 
 \Big)|u\rangle.
\end{split}
\end{equation*}


\subsection{New Quantum Fractional Transforms}\label{sec4sub2}

Having established the generalized dyadic-order framework, we now consider three concrete instantiations: the quantum fractional Hartley transform (QFrHT) and the fractional cosine-transform families associated with Types~I and~IV. 
To the best of our knowledge, explicit quantum-circuit realizations of these fractional constructions have not been previously reported in this form.

\begin{figure}[h]
\begin{center}
\includegraphics[width=13.4cm]{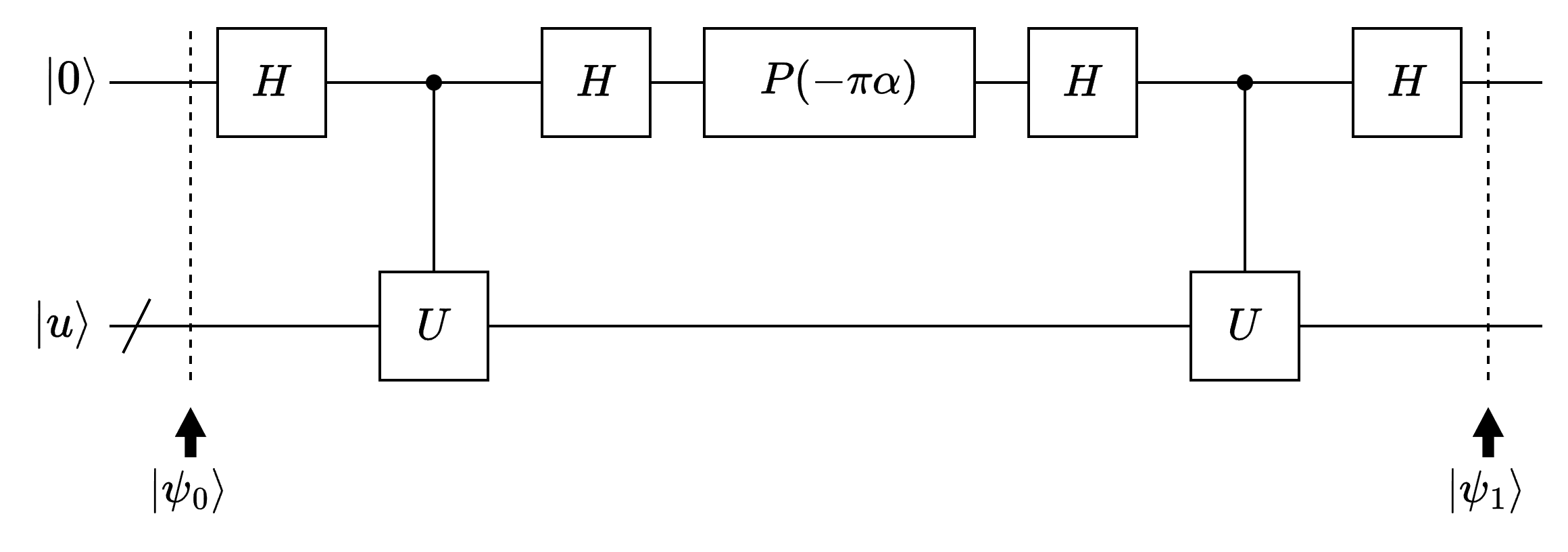}    
\caption{Diagram representing the quantum circuit for the fractional transform of an involution (QFrIn).} 
\label{fig:circuito_qfrt-2}
\end{center}
\end{figure}

Crucially, all three operators are involutions, meaning they possess an order of 2 (i.e., $U^2 = I$).
Consequently, their fractional realizations share a unified circuit architecture, as depicted in Figure~\ref{fig:circuito_qfrt-2}.
In this design, we exploit the fact that the single-qubit quantum Fourier transform is mathematically equivalent to the Hadamard gate. Therefore, we substitute the QFT block with $H$, simplifying the implementation using a standard quantum primitive.

The proposed circuit in Figure~\ref{fig:circuito_qfrt-2} mathematically realizes the operation
\begin{equation*} \label{eq:qfr_involution}
\begin{split}
    |\psi_1\rangle = QFrIn\{|\psi_0\rangle\} 
     = |0\rangle \Big[\frac{ (1+e^{-i\pi\alpha})}{2}I + \frac{(1-e^{-i\pi\alpha})}{2}U \Big]|u\rangle
\end{split}    
\end{equation*}

To obtain specific fractional transforms, the generic involutive operator $U$ is replaced by the corresponding circuit implementation of the QHT and of the cosine/sine transform structures discussed below.
Key properties of these base transforms are briefly outlined below, while a detailed 
derivation can be found in~\cite{Klappenecker:qct:qht:2001}.

\subsubsection{QHT}\label{sec4sub2sub1}
\begin{figure}[ht]
\begin{center}
\includegraphics[width=13.4cm]{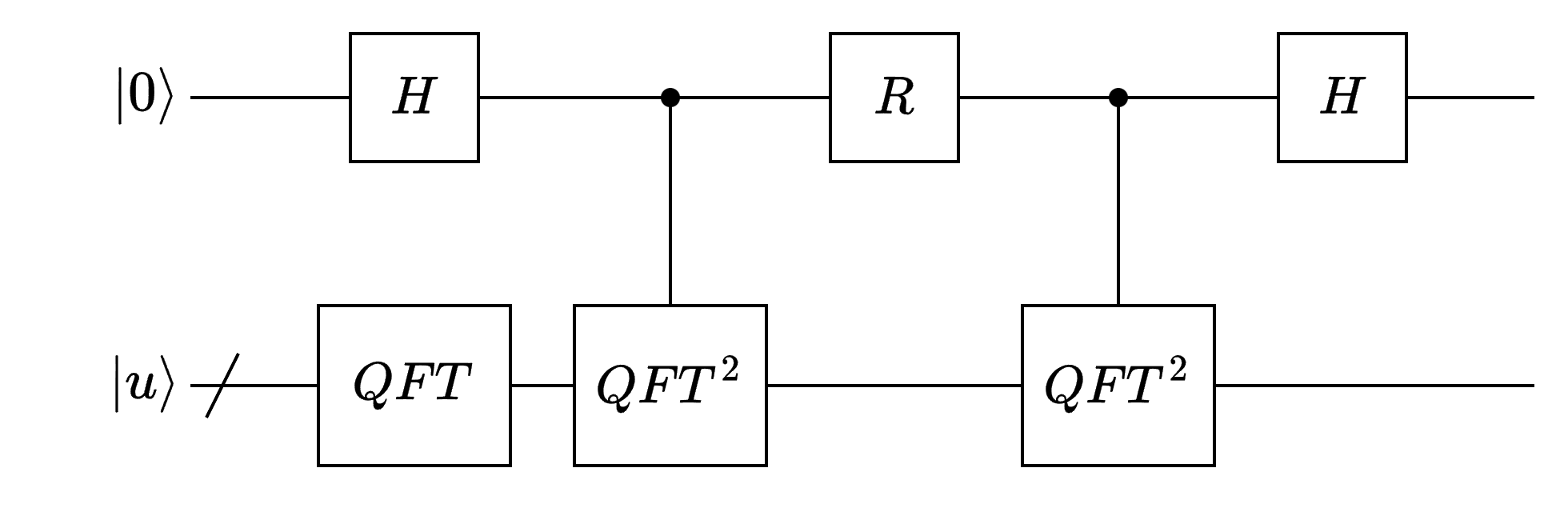}    
\caption{Diagram representing the quantum circuit for the QHT.}
\label{fig:circuito_qfrht}
\end{center}
\end{figure}

The circuit of the QHT makes use of an extra ancilla qubit, as shown in Figure \ref{fig:circuito_qfrht}.
%

The $R$ gate in the circuit has the form
\begin{equation*}
R = \frac{1}{2}\begin{bmatrix}
    1+i & 1-i \\
    1-i & 1+i
\end{bmatrix}.
\end{equation*}
This gate can be implemented as $HSH$, where $S=P(\pi/2)=\mathrm{diag}(1,i)$ is the phase gate.
An alternative implementation of the QHT can be seen at \cite{Wu_2024}.


\subsubsection{QCT-I}\label{sec4sub2sub2}

Following the construction of Klappenecker~\cite{Klappenecker:qct:qht:2001}, which in turn follows the approach of Wickerhauser~\cite{wickerhauser2019adaptedWaveletBook}, the Type-I cosine/sine circuit is obtained from a basis change applied to the DFT matrix of $2N$ points. 
The resulting operator has the block-diagonal form
\[
DCT_{N+1}^{I}\oplus DST_{N-1}^{I},
\]
with $N=2^n$, and the corresponding circuit acts on $n+1$ qubits.
%

\begin{figure}[h]
\begin{center}
\includegraphics[width=13.4cm]{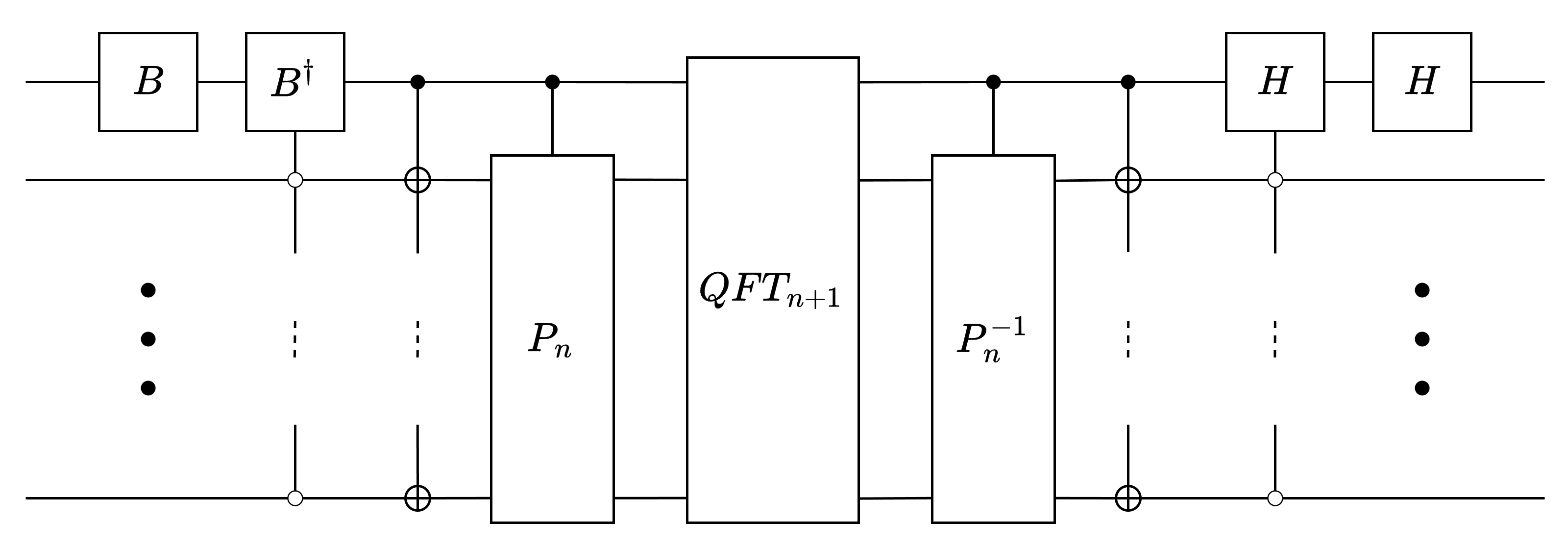}    
\caption{Diagram representing the quantum circuit for the QCT-I.} 
\label{fig:circuito_qfrct-i}
\end{center}
\end{figure}

The gates $B$ and $B^{\dagger}$ have the form
\begin{equation*}
\begin{split}
B = \frac{1}{\sqrt{2}} 
\begin{bmatrix}
    1 & i \\
    1 & -i
\end{bmatrix} = H \cdot S ; 
\quad 
B^{\dagger} = \frac{1}{\sqrt{2}} 
\begin{bmatrix}
    1 & 1 \\
    -i & i
\end{bmatrix} =  P(-\pi/2) \cdot H.
\end{split}
\end{equation*}

\begin{figure}[h]
\begin{center}
\includegraphics[width=8.4cm]{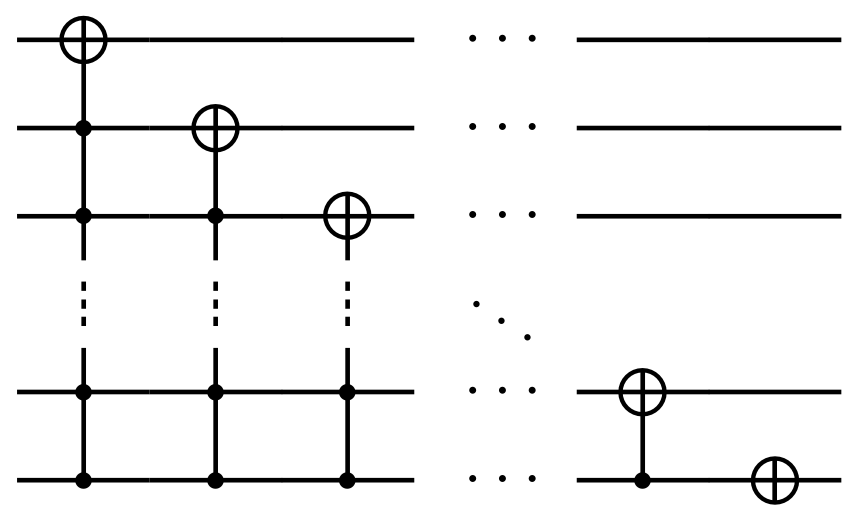}    
\caption{Diagram representing the quantum circuit for the $P_n$ gate~\cite{Klappenecker:qct:qht:2001}.} 
\label{fig:circuito_pn_gate}
\end{center}
\end{figure}

The $P_n$ gate is the shift operator $|x\rangle \rightarrow |x+1$ mod $2^n\rangle$ 
and it is implemented as shown in Figure \ref{fig:circuito_pn_gate}.

Since the implemented operator is the direct sum $DCT_{N+1}^{I}\oplus DST_{N-1}^{I}$, the cosine and sine components are not separated by a selector qubit. 
For this reason, this construction is more accurately described as a quantum cosine--sine transform of Type~I (QCST-I).
However, we will keep using its the usual name for compatibility with the literature.

\subsubsection{QCT-IV}\label{sec4sub2sub3}

\begin{figure}[h]
\begin{center}
\includegraphics[width=13.4cm]{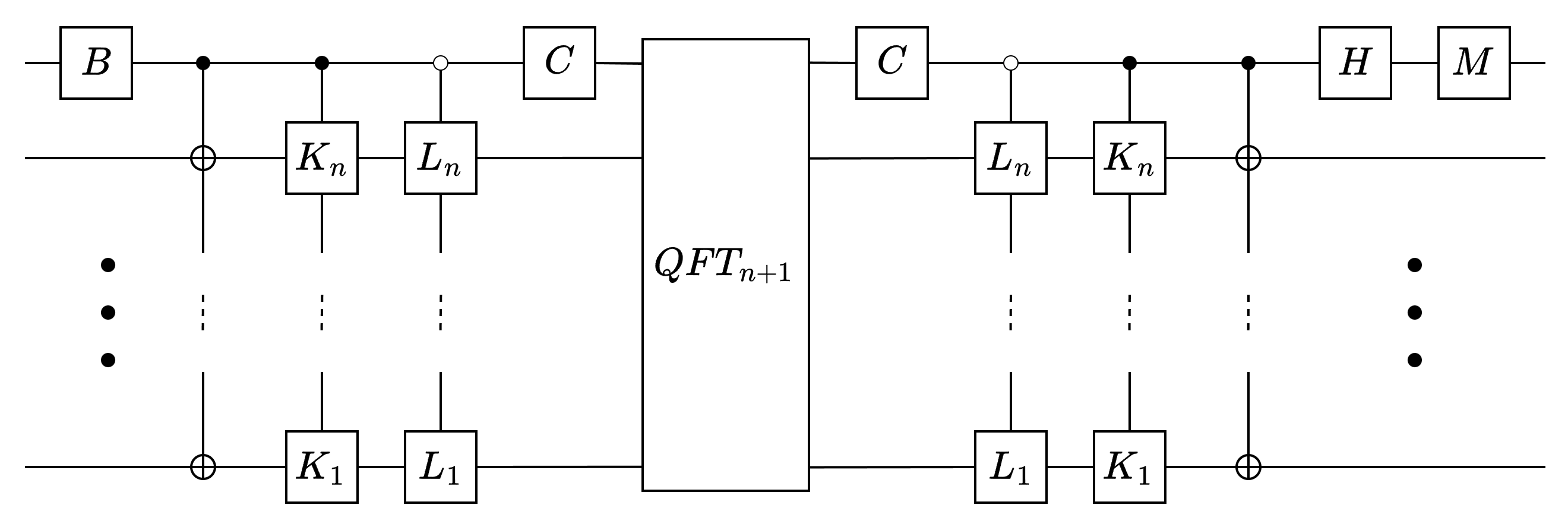}    
\caption{Diagram representing the quantum circuit for the QCT-IV.} 
\label{fig:circuito_qfrct-iv}
\end{center}
\end{figure}

By the same process of the QCT-I, Figure \ref{fig:circuito_qfrct-iv} shows the modified QCT-IV circuit with the operator matrix as $DCT_{N}^{IV} \oplus DST^{IV}_{N}$.

The new gates in the circuit have the following form: 
\begin{equation*}
\begin{split}
L_j &= \begin{bmatrix}
    1 & 0 \\
    0 & e^{i\pi 2^{j-1}/N}
\end{bmatrix}
= P\!\left(\frac{\pi 2^{j-1}}{N}\right), \\
K_j &= X \cdot L_j \cdot X, \\
C &= \begin{bmatrix}
    1 & 0 \\
    0 & e^{i\pi/2N}
\end{bmatrix}
= P\!\left(\frac{\pi}{2N}\right), \\
M &= e^{-i\pi/4N} I,
\end{split}
\end{equation*}
where $j=1,2,\dots,n$ and $P(\phi)=\mathrm{diag}(1,e^{i\phi})$.

Unlike the Type-I case, the operator $DCT_{N}^{IV}\oplus DST_{N}^{IV}$ admits a direct selector-qubit interpretation:
\begin{equation*}
\begin{split}
    QCT^{IV}_{n+1}\{|0u\rangle\} = |0\rangle C^{IV}_n|u\rangle; 
    \quad 
    QCT^{IV}_{n+1}\{|1u\rangle\} = |1\rangle S^{IV}_n|u\rangle,
\end{split}    
\end{equation*}
where $C^{IV}_n = DCT^{IV}_N$ and $S^{IV}_n = DST^{IV}_N$.

\section{Conclusion}\label{sec:conclusion}

In this paper, building upon the circuit architecture of the quantum fractional Fourier transform (QFrFT), we presented a generalized circuit framework for constructing Shih-type fractionalizations of unitary operators of dyadic order, i.e., operators $U$ satisfying $U^{2^n}=I$. 
The proposed construction realizes the fractional operator as a coherent linear combination of the integer powers of $U$, generated through ancilla-domain Fourier processing and diagonal phase modulation.
As concrete applications, we instantiated the framework for the quantum Hartley transform (QHT) and for the cosine/sine transform structures associated with Types~I and~IV, obtaining explicit circuit realizations of the corresponding fractional transforms.
To the best of our knowledge, these constitute the first explicit quantum-circuit constructions of these fractional transform structures within the present dyadic-order framework.

In future work, we intend to investigate applications of these transforms in quantum signal-processing settings, particularly in areas where their classical counterparts have proved useful, such as watermarking, image encryption, and related processing tasks.

\section*{Acknowledgment}
The authors acknowledge the QUANTA Research Group (Department of Physics, UFPE) for the resources provided for this work.

\bibliography{IEEEtran}

\begin{IEEEbiography}[{\includegraphics[width=1in,height=1.25in,clip,keepaspectratio]{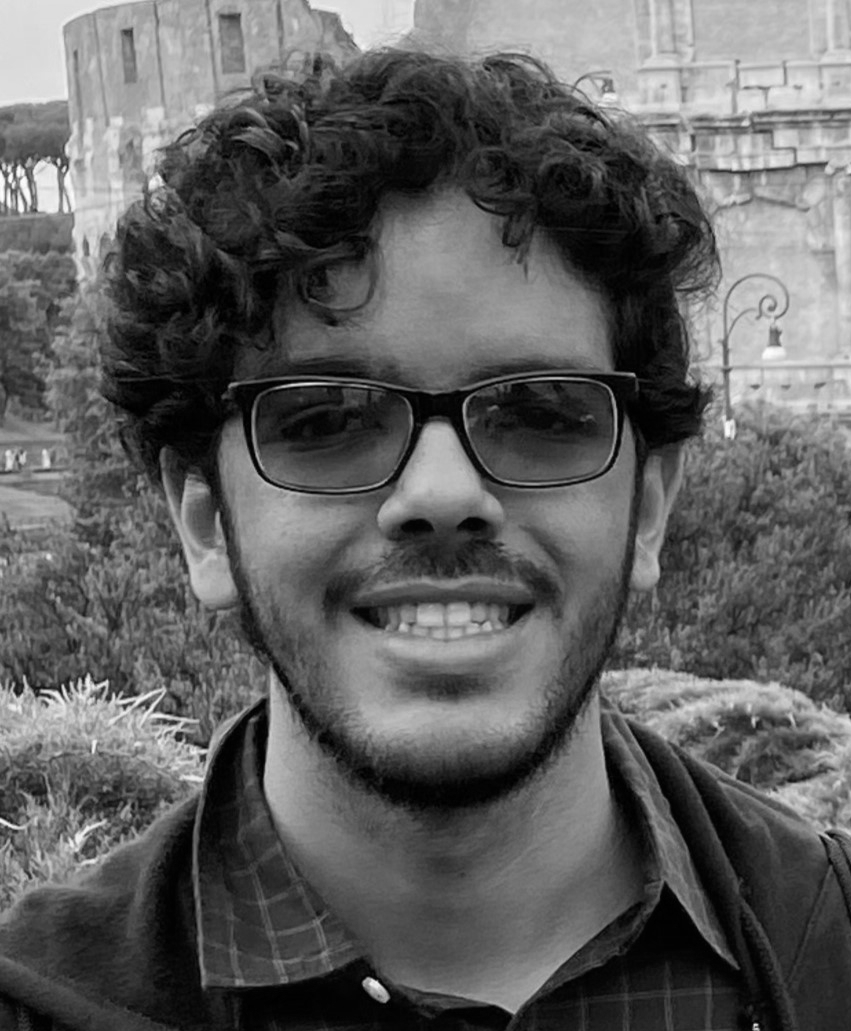}}]{MATHEUS J. A. OLIVEIRA} was born in Brazil. He received the bachelor’s degree in electronic engineering from the Universidade Federal de Pernambuco, Recife, Brazil, in 2024, where he is currently pursuing the M.Sc. degree in electrical engineering and is engaged in quantum computing research, in particular quantum information processing.
\end{IEEEbiography}

\begin{IEEEbiography}[{\includegraphics[width=1in,height=1.25in,clip,keepaspectratio]{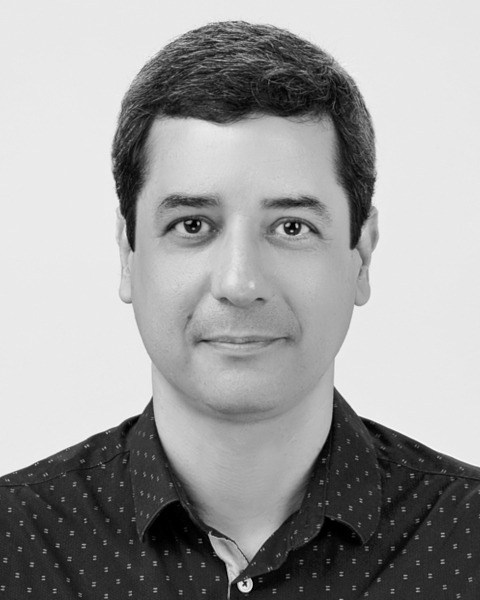}}]{ISRAEL F. ARAUJO}
received the M.Sc. degree in Physics from the Universidade Federal Rural de Pernambuco, Brazil, and the Ph.D. degree in Computer Science from the Universidade Federal de Pernambuco, Brazil. He subsequently completed postdoctoral research at Yonsei University, Republic of Korea, and at the Department of Electronics and Systems, Universidade Federal de Pernambuco, Brazil. He is currently a Lead Quantum Scientist at data cybernetics ssc GmbH.
\end{IEEEbiography}

\begin{IEEEbiography}[{\includegraphics[width=1in,height=1.25in,clip,keepaspectratio]{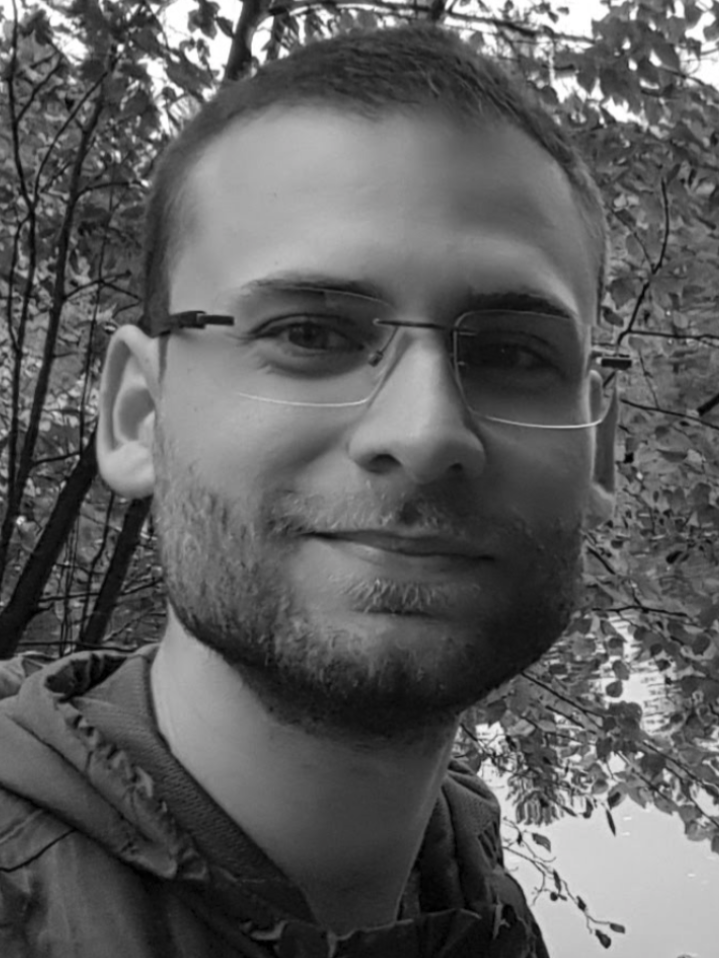}}]{JOS\'E R. de OLIVEIRA NETO} was born in Brazil in 1989. He received the B.Sc., M.Sc., and Ph.D. degrees in electrical engineering from the Federal University of Pernambuco (UFPE), Brazil, in 2013, 2015, and 2019, respectively. He is currently an Assistant Professor with the Department of Electronics and Systems, UFPE. His research interests include digital signal processing, embedded systems, and hardware implementations.
\end{IEEEbiography}

\begin{IEEEbiography}[{\includegraphics[width=1in,height=1.25in,clip,keepaspectratio]{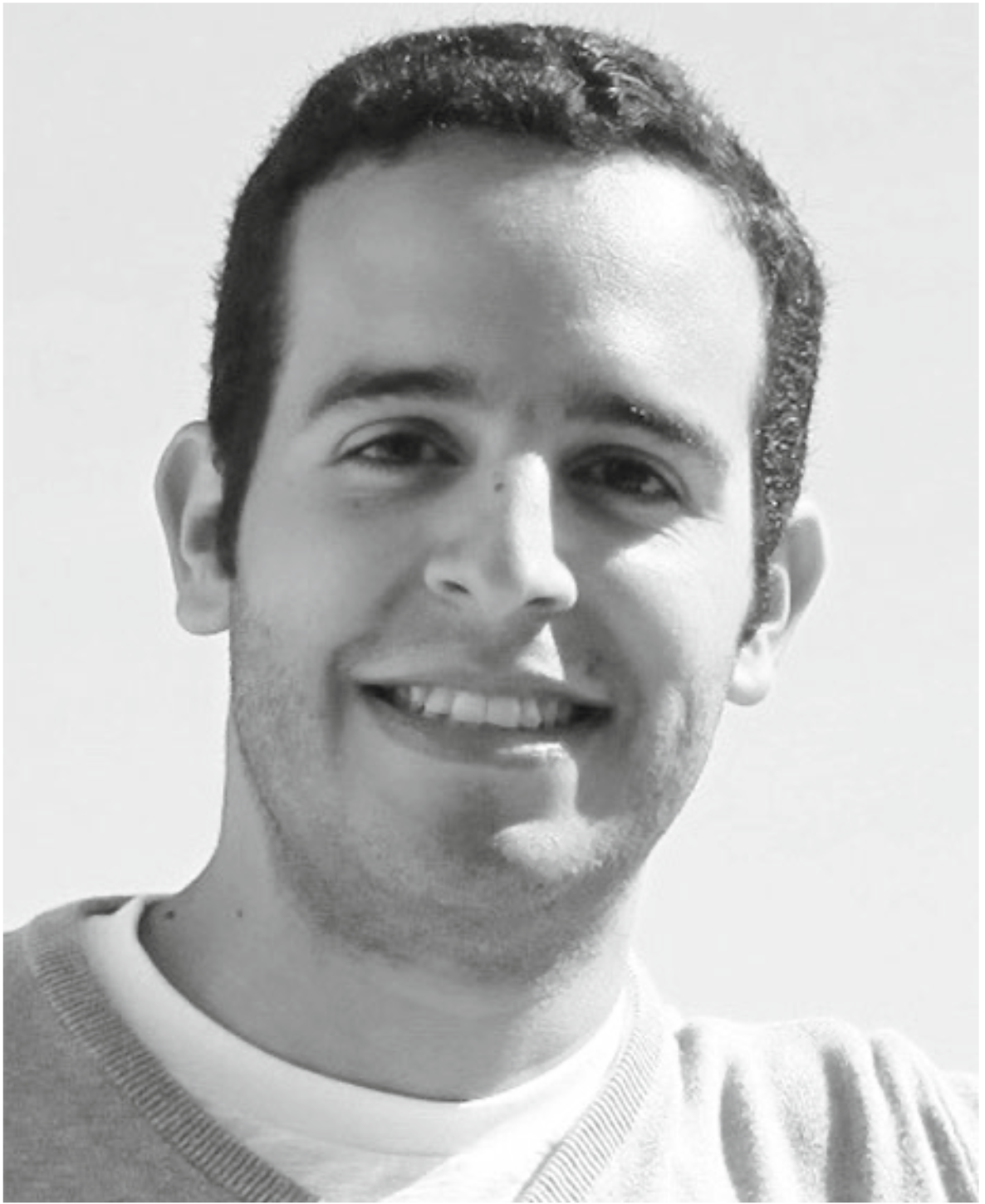}}]{JULIANO B. LIMA} was born in Brazil where he studied electrical engineering. He received the M.Sc. and Ph.D. degrees in electrical engineering from Federal University of Pernambuco (UFPE), Brazil, in 2004 and 2008, respectively. Since 2015, he has been a research productivity fellow awarded by the Conselho Nacional de Desenvolvimento Cient\'ifico e Tecnol\'ogico. He is currently an Associate Professor of the Department of Electronics and Systems at UFPE. His main research interests are related to signal processing theory and its applications in time series analysis, graph-structured data, neuroscience, and cryptography.
\end{IEEEbiography}

\end{document}